\documentclass[aps,prb,final,reprint,twocolumn,floatfix,showpacs,superscriptaddress,notitlepage, longbibliography]{revtex4-2}

\usepackage[latin9]{inputenc}
\usepackage{graphicx,amssymb,soul,color,amsmath,bm}
\usepackage{epstopdf,hyperref}
\usepackage{braket,dsfont}
\usepackage[normalem]{ulem}

\usepackage{graphicx}
\usepackage{subfigure}

\hypersetup{colorlinks=true,citecolor=blue,linkcolor=magenta}

\sethlcolor{yellow}
\usepackage{xcolor}

\newcommand{\adrian}[1]{\textcolor{blue}{#1}}

\begin{document}

\title{
% Quasi-Fermi liquid behavior in one-dimensional interacting quantum fluids \\
\adrian{Quasi-Fermi liquid behavior in a one-dimensional system of interacting spinless fermions}
}

\author{Joshua D. Baktay}
\affiliation{Department of Physics, Northeastern University, Boston, Massachusetts 02115, USA}

\author{Alexander V. Rozhkov}
\affiliation{Institute for Theoretical and Applied Electrodynamics, Russian Academy of Sciences, 125412 Moscow, Russia}

\author{Adrian E. Feiguin}
\affiliation{Department of Physics, Northeastern University, Boston, Massachusetts 02115, USA}

\author{Juli\'an Rinc\'on}
\affiliation{Department of Physics, Universidad de los Andes, Bogot\'a D.C. 111711, Colombia}

\date{\today}

\begin{abstract}
We present numerical evidence for a new paradigm in one-dimensional interacting fermion systems, whose phenomenology has traits of both, Luttinger liquids and Fermi liquids. This new state, dubbed a \emph{quasi-Fermi liquid}, possesses a discontinuity in its fermion occupation number at the Fermi momentum. The excitation spectrum presents particle-like quasiparticles, and absence of hole-like quasiparticles, giving rise instead to edge singularities. Such a state is realized in a one-dimensional spinless fermion lattice Hamiltonian by fine-tuning the interactions to a regime where they become irrelevant in the renormalization group sense. We show, using uniform infinite matrix products states and finite-entanglement scaling analysis, that the system ground state is characterized by a Luttinger parameter $K = 1$ and a discontinuous jump in the fermion occupation number. We support the characterization with calculations of the spectral function, that show a particle-hole asymmetry reflected in the existence of well-defined Landau quasiparticles above the Fermi level, and edge singularities without the associated quasiparticles below. These results indicate that the quasi-Fermi liquid paradigm can be realized beyond the low-energy perturbative realm. %Finally, we explore other models likely to realize this state, instantiating it as a potentially new universality class.
\end{abstract}

\maketitle

\newpage

\section{Introduction}
It is well understood that a fairly generic, interacting quantum system in one spatial dimension (1D) can be described at low energies by Luttinger liquid (LL) theory~\cite{haldane1981}. This proposal overrules the standard Fermi liquid (FL) theory due to the reduced phase-space scattering between particles, preventing the formation of quasiparticles and forcing instead the existence of collective excitations~\cite{gogolinbook, giamarchi}. The conventional LL paradigm involves linearizing the dispersion relation around the Fermi surface, which consists of two points. This approximation is justified because the band curvature is irrelevant in the renormalization group sense for energy scales arbitrarily close to the Fermi level. Indeed, spectral function calculations are asymptotically exact but only in the low-energy, perturbative regime~\cite{Dzya1974, Luther1974, Meden1992, Voit1993, Schonhammer1993, Imambekov2009, Imambekov2009b, Pereira2009}. However, capturing more realistic physics requires accounting for irrelevant interactions at higher energy scales where linearization is no longer appropriate~\cite{Pustilnik2006, Rozhkov2006, Khodas2007, Periera2008, Imambekov2009, Imambekov2009b, Pereira2009, Imambekov2012, Periera2012, Rozhkov, Essler2015}.

Attempting to apply LL theory at high energies unavoidably leads to modifications, giving rise to unexpected behavior that defies common knowledge. Phenomenology beyond this ``linear'' LL paradigm include considering corrections to the theory around a stable fixed point that can be associated to (i) the curvature of the dispersion relation, (ii) irrelevant interactions, or (iii) a momentum-dependent interaction potential~\cite{Pereira2009, Imambekov2009, Imambekov2009b, Imambekov2012, Markhof2016, Meden1999, Schonhammer1993, Periera2012}.

In a FL, irrelevant terms can be accounted for perturbatively. They broaden the spectral function, maintaining the quasiparticle picture. This perturbation-theory approach fails in LL theory. For example, at finite dispersion curvature, the Lorentz invariance and particle-hole symmetry introduced by the linearization are both broken, leading to on-shell divergences~\cite{Periera2012, Imambekov2012}. Together, these symmetry breaking effects lift the degeneracy in the dispersion relation leading to qualitatively new behavior in the dynamic correlations~\cite{Pustilnik2006, Periera2008}. The differences are twofold. (1) Near the Fermi momentum, $k_F$, the spectral functions of both the linear and nonlinear LLs possess power-law singularities but with different scaling exponents~\cite{Periera2012, Imambekov2012}. (2) Away from $k_F$, the excitation spectrum can develop finite lifetime quasiparticles or power-law excitations~\cite{Pereira2009, Imambekov2012, Markhof2016, Essler2015}.

We present a microscopic, spinless fermion model that, contrary to the conventional paradigms, simultaneously exhibits both FL and LL characteristics. This \textit{quasi-Fermi liquid} (qFL) was originally presented in Refs.~\onlinecite{Rozhkov2006, Rozhkov} in the context of a 1D continuum Hamiltonian that contains only irrelevant interactions, with scaling dimension three. The model was found to realize a discontinuity in the momentum distribution while lacking the perturbatively defined fermionic quasiparticles normally indicated by such a discontinuity. The former hints to FL behavior while the latter indicates LL behavior. We numerically study the ground state of the equivalent lattice problem directly in the thermodynamic limit using uniform matrix product states (uMPS)~\cite{Vidal2007, Haegeman2011, schollwock2011, lecture_notes} and the variational uMPS (VUMPS) algorithm~\cite{VUMPSalg, lecture_notes}. Our VUMPS results are further refined using finite-entanglement scaling (FES)~\cite{Tagliacozzo2008, FES, Cincio2018, FES2}. Evidence for the qFL is further strengthened by characterizing the excitation spectrum using large-scale time-dependent density matrix renormalization group (tDMRG) calculations~\cite{White2004a, Daley2004, vietri, Paeckel2019}.

The main goal of this work is to show, using uMPS, VUMPS, FES, and tDMRG, that the universality class defined by the qFL paradigm can be detected in the nonperturbative regime in quantum lattices. This is accomplished by analyzing the ground and excited states of a lattice Hamiltonian at different fillings such that the ground state exhibits FL behavior, while the excitation spectrum suggests similarities with the non-linear LL. %--namely asymmetric behavior in the particle and hole sectors of the spectral functions. 
This physics is realized along a critical line with Luttinger parameter $K=1$ in Hamiltonian parameter space, legitimizing the departure from the standard LL and FL paradigms of quantum fluids.

%that the universality class defined by the qFL paradigm goes beyond the low-energy perturbative realm and, in fact, its phenomenology can be detected in the non-perturbative regime. In other words, we show that there exist 1D lattice Hamiltonians with a phase diagram that displays qFL behavior, thereby departing from the standard LL and FL paradigms of quantum fluids.

The article below has the following structure: In Sec.~\ref{sec:model}, we introduce our model and the relevant numerical methods. In Sec.~\ref{sec:results}, the momentum distribution and the charge static structure factor are computed and analyzed for evidence of both FL and LL behavior. This is followed by calculations of the spectral function in both the particle and hole sectors, as well as a scaling analysis of various momentum cuts. We close with a discussion of the results and their possible implications in Sec.~\ref{sec:conclusions}.

\section{Model and methods\label{sec:model}}

In this work we focus on a 1D, extended, spinless fermion model on the lattice with interactions to first and second neighbors. The  Hamiltonian is written as:
%$H = \sum_{j \in \mathbb{Z}} h_{j}$, where the Hamiltonian local terms are given by
\begin{equation}
\begin{aligned}
H = &-t \sum_j \bigl( c_j^\dagger c_{j + 1} + \textrm{H.c.} \bigr) + \mu \sum_j \tilde n_j \\
&+ V\sum_j  \tilde n_j \tilde n_{j + 1} + V_2 \sum_j \tilde n_j \tilde n_{j + 2}, 
\end{aligned}
\end{equation}
with $\tilde n_j = c^\dagger_jc_j - 1/2$ giving the particle-hole symmetric version at $\mu = 0$. Parameters $V$ and $V_2$ are the nearest-neighbor and next-nearest-neighbor interactions, respectively, and $\mu$ is the chemical potential; all energies are expressed in units of the hopping $t$. 

{The phase diagram in the $V > 0,\, V_2 > 0$ region was characterized in Ref.~\onlinecite{Mishra} at half filling using finite-size DMRG. Four phases were identified: two charge-density-wave insulating phases, a LL phase, and a bond-order phase. The charge-density-wave regions arise when either $V$ or $V_2$ ($V'$ in their notation) is appreciably greater than the other. In between these two regions, where $V$ and $V_2$ are comparable, there is a LL phase that transitions into the bond-order phase at higher values of $V$ and $V_2$.}

According to Ref.~\onlinecite{Rozhkov}, for a generic system of interacting fermions, the qFL state can be stabilized by fine-tuning the interaction couplings such that marginal interactions are nullified. The influence of relevant interactions such as umklapp, which occur at commensurate fillings, can be avoided by setting $\mu \neq 0$. The remaining \emph{irrelevant} interactions stabilize this unique state. In our case, this involves tuning $V$ and $V_2$ with the constraint $V \cdot V_2 < 0$; {\it i.e.} if one is attractive, the other is repulsive.

%\section{Numerical Methods}

In order to account properly for the singular behavior of the 1D problem, it is necessary to resort to methods that work directly in the thermodynamic limit~\cite{Karrasch2012}. To this aim, the translationally invariant ground state is represented by a uMPS~\cite{Vidal2007, Haegeman2011}, characterized by its bond dimension, $\chi$, which controls the size of its matrices and defines a variational manifold as a subspace of the exponentially large, many-body Hilbert space. The minimum within this manifold, with respect to the cost function $\braket{{\rm uMPS}|H|{\rm uMPS}}$, corresponds to the approximate ground state. It is reached via the DMRG-inspired VUMPS algorithm~\cite{VUMPSalg}, which iteratively finds the ground state and its energy, subject to an error threshold criterion. Convergence to a variational minimum is characterized by the energy density error and the norm of the tangent vector. Our simulations achieved energy density errors and tangent vector norms on the order of $10^{-12}$ and $10^{-13}$, respectively. The quality of our uMPS approximation to the exact ground state wavefunction is captured by the discarded weight as a function of $\chi$. Our simulations included bond dimensions up to $\chi = 640$ which yield discarded weights of the order of $10^{-10}$. For more details on the VUMPS algorithm we refer the reader to Refs.~\onlinecite{VUMPSalg, lecture_notes}.

By virtue of the translational invariance of our system, the uMPS ansatz works directly in the thermodynamic limit, thus removing any finite-size effects. However, uMPS are still approximate, though quasi-exact, variational wavefunctions due to the finite bond dimension $\chi$. Physically, $\chi$ captures the approximate amount of entanglement present in the system, and can be tuned to improve accuracy. A finite $\chi$ introduces finite-entanglement effects analogous to finite-size effects. This can be overcome using a FES analysis, which allows one to extrapolate physical quantities of interest to the infinite-bond dimension limit, by analyzing its behavior against a suitable length scale parameter~\cite{Tagliacozzo2008, FES, Cincio2018, FES2}. This FES analysis was done for all relevant expectation values to follow.

\section{Results\label{sec:results}}

\subsection{Ground state}

The general characteristics of the FL can be understood in terms of the Landau conjecture which states that the ground state and low-energy excited states of a FL are adiabatically connected to the free Fermi gas, exhibiting nonzero overlap between them. On the other hand, the ground and excited states of a LL have zero overlap with the noninteracting case, thus constituting their own universality class. For the qFL, because its usually dominant, marginal interactions have been nullified, one expects its ground state to be perturbatively connected to the ground state of the free Fermi gas. However, in the infrared limit, the remaining, technically irrelevant interactions now have nontrivial effects on its excited states; they have no fermionic quasiparticles, which implies no overlap with the noninteracting case, like a LL. (Indeed, one might instead expect bosonic excitations; see discussion below.) %Thus, ``half'' of Landau's conjecture is satisfied.

\begin{figure}
    \includegraphics[width=.43\textwidth]{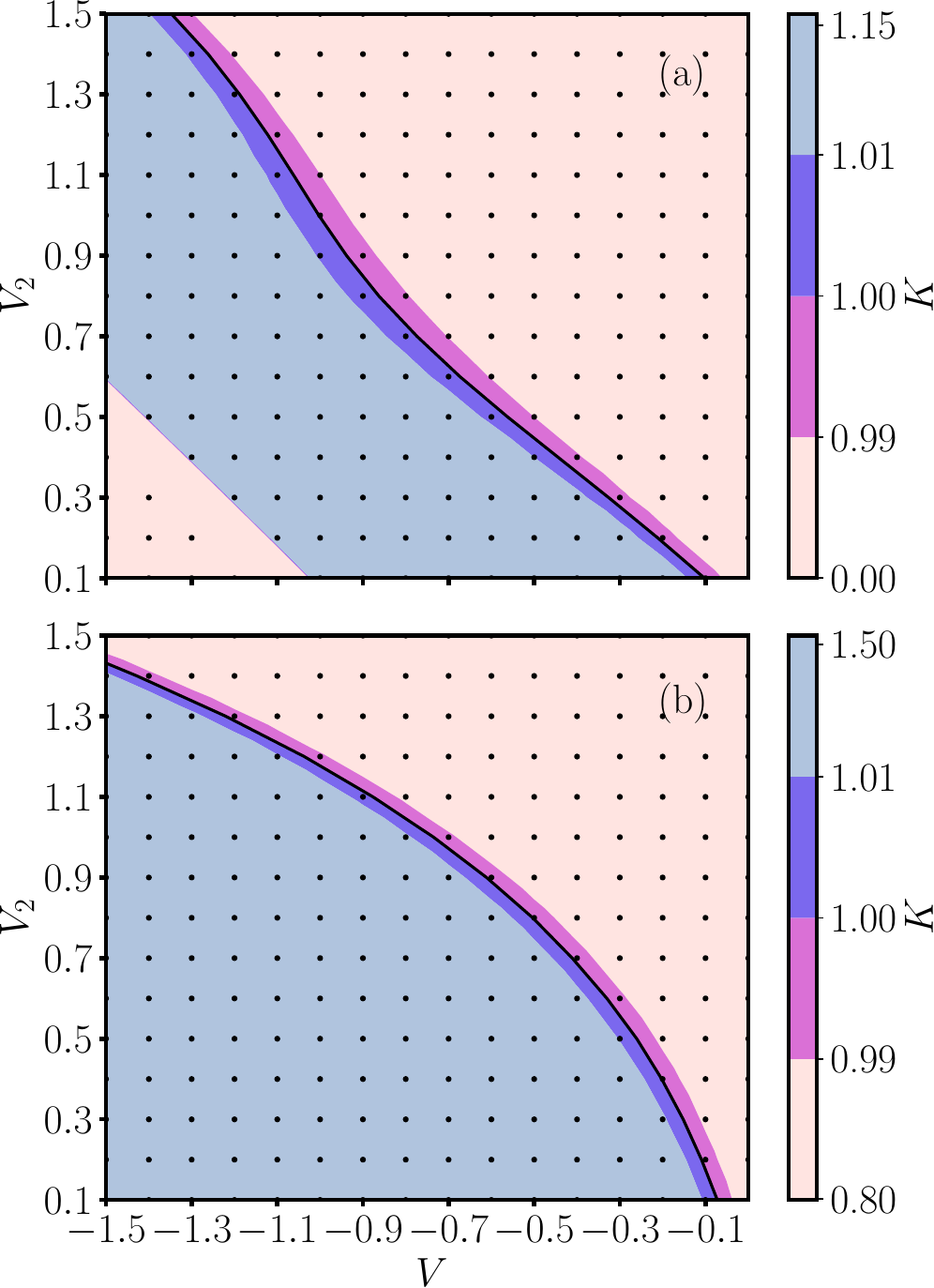}
    \caption{Phase diagram of Luttinger parameter, $K$, for $V < 0$ and $V_2 > 0$ at $\mu/t = +1.0$ \label{fig:Kphase}(a) and $\mu/t = +0.5$ (b). The quasi-Fermi liquid state is stable along the black curve which is our best estimate of the critical line, obtained as an interpolation of the data points within the band of qFL candidates, $0.99 < K < 1.01$. The band can be made arbitrarily precise with a step size $0 < \delta \leqslant 0.1$ Above this band the ground state is dominated by charge-ordered fluctuations. Below the band, the ground state is dominated by superconducting fluctuations~\cite{giamarchi}. For $\mu/t = +1.0$ (a) in the lower left corner, as $V$ becomes much greater than $V_2$ the particles are driven into a very sparse, charge-ordered state. This breaks translation invariance which prevents ground state convergence resulting in the missing data points. Those that converge have $K \approx 0$.
    \label{fig:phasediag}
    }
\end{figure}

According to the above argument, it is then natural to seek ground states that have a Luttinger parameter $K = 1$. Such parameter accounts for the nature and strength of the interaction in the LL. It follows from LL theory that $K$ can be extracted as a low-momentum, linear approximation of the charge structure factor (CSF), the Fourier transform of the density-density correlation function, 
\begin{equation}
    D(k)=\sum_{j}\langle n_0 n_j\rangle e^{-ikj}.
\end{equation}
The CSF was computed for the uMPS representing the ground state in the thermodynamic limit~\cite{lecture_notes}. For known $D(k)$, the parameter $K$ can be extracted from the slope near $k \to 0^+$: $dD(k)/dk=K/\pi$~\cite{gogolinbook, giamarchi, Daul1998}. A grid search was carried out through a wide range of the parameter space, depicted in Fig.~\ref{fig:phasediag}, where a narrow band is seen whose values of $V$ and $V_2$ yield $K \approx 1$. Crucially, these values of $V, V_2$ are generally comparable in magnitude but not perfectly equal, highlighting the importance of a judicious grid search. The pairs $(V, V_2)$, with fixed $\mu$, represent a family of qFL candidates for further investigation. 

\begin{figure}
    \includegraphics[width=.48\textwidth]{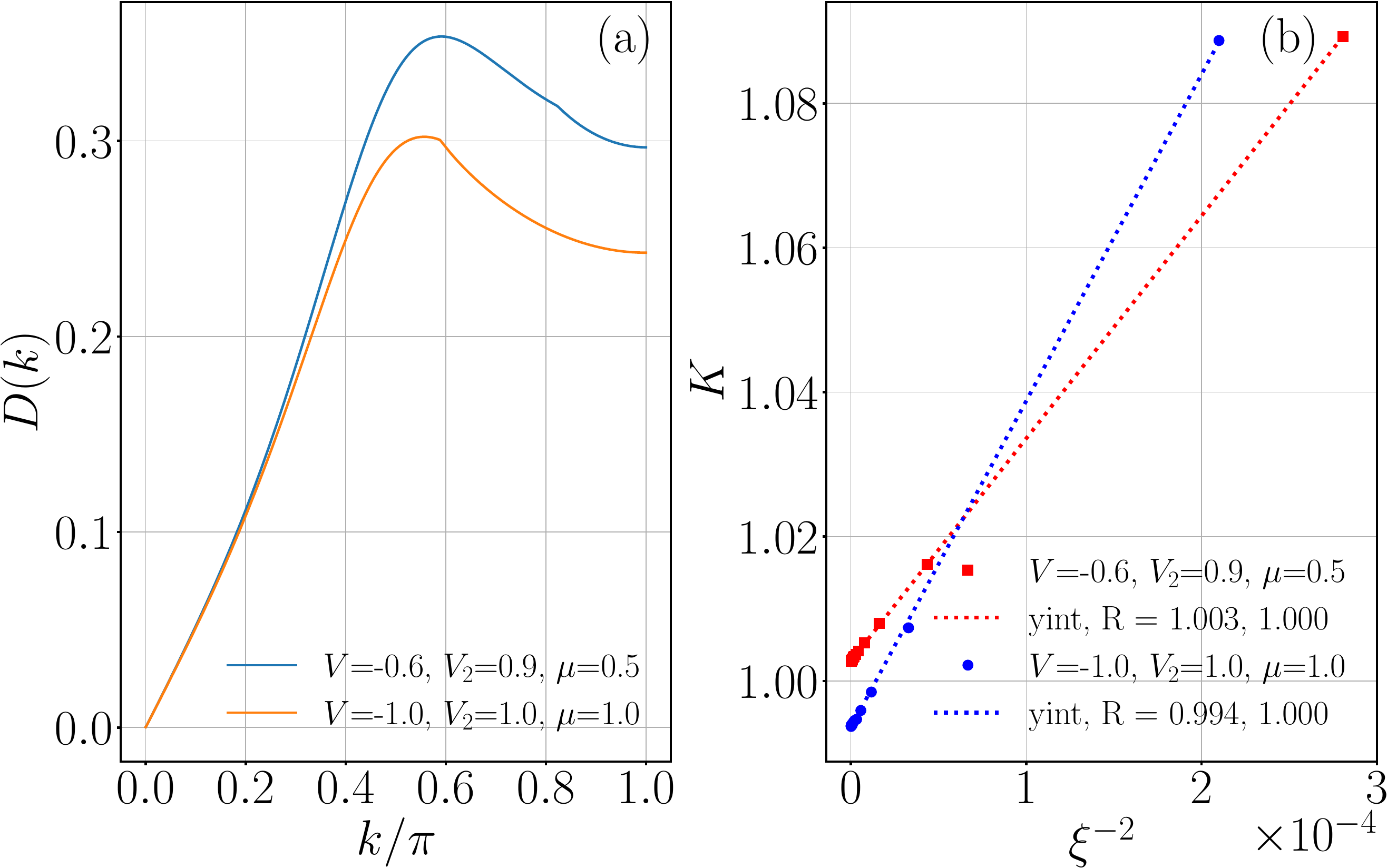}
    \caption{Static charge structure factor $D(k)$ vs $k$ (a) and the Luttinger parameter $K$ vs $\xi^{-2}$ (b) for two quasi-Fermi liquid candidates. Panel (a) contains a family of $D(k)$ curves, plotted for different values of $\chi$, for each candidate. Individual curves in each family are not discernible due to the lack of change as $\chi$ increases (see main text). Panel (b) shows $K$ in the infinite $\chi$ limit as the intercept of a linear fit. Our two candidates converge to $K = 1.0028$ and $0.9936$ with statistical correlation $\simeq 1$.
    \label{fig:CSF}
    }
\end{figure}

In order to thoroughly characterize the state of matter emerging along the $K = 1$ line, as shown in Fig.~\ref{fig:phasediag}, we now present a full analysis for two sets of representative parameter value candidates: $(V, V_2, \mu) \in \{(-0.6, 0.9, 0.5), (-1.0, 1.0, 1.0)\}$~\cite{suppmat}.

The CSF for the two candidates we highlight can be seen in Fig.~\ref{fig:CSF}(a). The deviations at high momenta from the plateau-like trend of free fermions are a signature of interacting behavior. 
%Also notice that the CSF does not change significantly with $\chi$. 
The overall smoothness of the curves indicates the absence of charge order. We point out that correlation functions can change significantly with $\chi$ (because of its approximate nature), revealing spikes in the CSF at higher $\chi$ in the presence of charge order. Here, we do not observe this behavior and instead conclude that the ground state is metallic.  

We conducted a FES analysis to determine $K$ for our candidates in the infinite-bond dimension limit, based on finite-bond dimension data. This is displayed in Fig.~\ref{fig:CSF}(b), where $K$ is plotted versus the correlation length of the MPS, $\xi := \xi_1(\chi)$, defined as 
\begin{equation}
\xi_1(\chi)^{-1} = -\log|\lambda_{1}(\chi)|,
\end{equation}
where $\lambda_{1}(\chi)$ is the second largest eigenvalue of the MPS transfer matrix~\cite{Tagliacozzo2008, FES, FES2, Cincio2018, lecture_notes}. A similar analysis can be done using the length scale $\delta = \xi_n^{-1} - \xi_1^{-1}$, in place of $\xi_1(\chi)$, where $\xi_n^{-1} = -\log|\lambda_{n}(\chi)|$ and $\lambda_{n}(\chi)$ is the $n$-th sub-leading eigenvalue of the MPS transfer matrix~\cite{Cincio2018}; other combinations of $\xi_n$'s are also possible~\cite{FES2}. Indeed, we have found an analogous scaling behavior to that shown in Fig.~\ref{fig:CSF}(b) for two equivalent definitions: $\delta = \xi_2^{-1} - \xi_1^{-1}$ and $\delta = \xi_3^{-1} - \xi_1^{-1}$ (not shown)~\cite{suppmat}. The length scales $\xi$ and $\delta$ control the finite-entanglement scaling properties of the MPS approximation. It is expected that as $\chi \to \infty$, both $\xi \to \infty$ and $\delta \to 0$. A linear fit yields an intercept value equal to the extrapolated value of $K$ for $\xi \to \infty$. Given the high statistical correlation of the fit, we can be confident in the value of $K$ and the legitimacy of the $K \approx 1$ band in parameter space.

Next, the momentum distribution function $n(k)$,
%in order to determine the presence of a discontinuity at the Fermi level,
\begin{equation}
    n(k)=\sum_{j}\langle c^\dagger_0 c_j \rangle e^{-ikj},
\end{equation}
was calculated similarly to the CSF~\cite{lecture_notes}. According to LL theory, $n(k)$ possesses a power-law singularity near the Fermi momentum $k_F$~\cite{haldane1981, gogolinbook, giamarchi}, 
\begin{equation}
n(k) \sim \mathrm{sgn}(k-k_F)\left|k-k_F\right|^{(K + K^{-1})/2-1},
\end{equation}
in sharp contrast to the discontinuity present in FL. %We believe that discriminating between a discontinuity and a singularity is sufficient to show whether our qFL ground state can be characterized by LL or FL behavior.

\begin{figure}
    \includegraphics*[width=.45\textwidth]{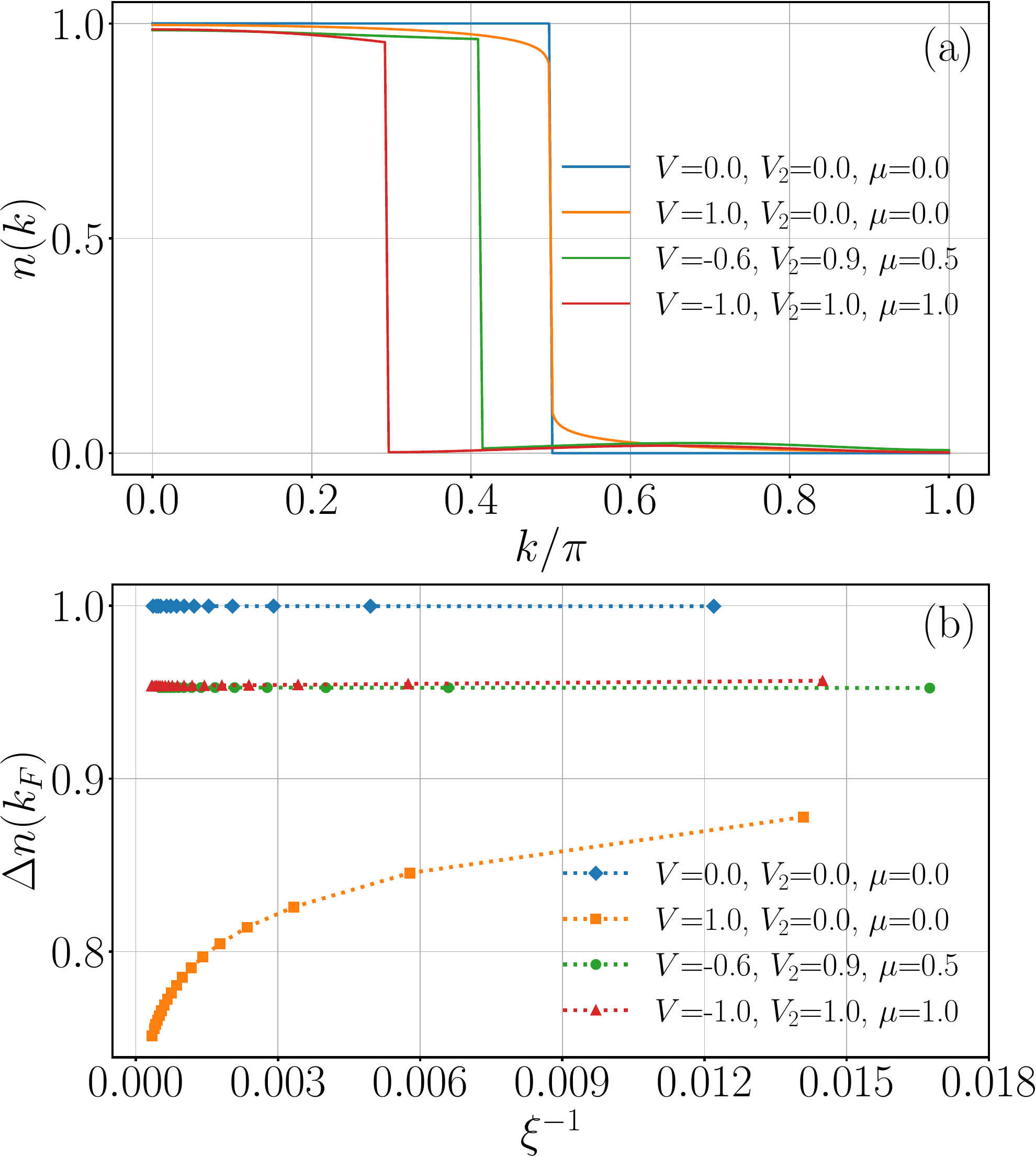}
    \caption{Momentum distribution $n(k)$ vs $k$ (a) and size of the discontinuity $\Delta n(k_{F})$ vs $\xi^{-1}$ (b) for both candidates compared to the free Fermi gas $(V, V_{2}) = (0, 0)$ (blue) and a known LL state $(V, V_{2}) = (1, 0)$ (yellow); clearly in panel (b) the scaling behavior of the LL differs greatly from the rest. The quasi-Fermi liquid candidates and the Fermi gas scale at most linearly, showing variations of order $10^{-3}$ whose intercept corresponds to $\Delta n(k_{F})$ as $\chi \to \infty$. The Fermi gas converges to 1, as expected, while our candidates converge to $\approx 0.95$, thus demonstrating the presence of a true discontinuity. The LL scales logarithmically, implying the existence of a singularity at $k_F$. This is the expected power-law behavior, converging to zero for $\xi \to \infty$.
    \label{fig:momfes}
    }
\end{figure}

A finite bond dimension introduces an inherent length scale via the MPS correlation length, $\xi$, that limits the resolution of all correlation functions and prevents one from easily discriminating between a singularity and a discontinuity in a finite-entanglement data set. Furthermore, extrapolating the numerical derivative at the $k_F$ (using FES) cannot differentiate between FL and LL as the derivative of the LL power law also diverges at the Fermi momentum~\cite{Karrasch2012}. To overcome this, a FES analysis was performed on the size of the discontinuity about $k_F$ computed as 
\begin{equation}
\Delta n(k_{F}) \equiv n(k_{F} - \pi / \xi (\chi) )  - n(k_{F} + \pi / \xi (\chi) )
\end{equation}
with $\xi(\chi)$ defined above. The momentum distribution functions for our qFL candidates can be seen in Fig.~\ref{fig:momfes}(a). Discontinuity $\Delta n(k_{F})$ versus correlation length can be seen in Fig.~\ref{fig:momfes}(b) compared against the Fermi gas and a known prototypical LL. Clearly the extrapolated behavior greatly differs between the two; the LL possesses a logarithmic scaling such that $\Delta n(k_{F}) \rightarrow 0$ as the correlation length increases, while the qFL displays at most a linear scaling such that $\Delta n(k_{F})$ approaches a nonzero constant as $\xi(\chi) \to \infty$. This behavior matches that of the free Fermi gas, confirming our analysis. Other quantities such as the entanglement entropy of the ground state can also be readily obtained~\cite{suppmat}.

The jump in $n(k)$ together with $K = 1$, as $\xi \to \infty$, corroborates the FL behavior of our qFL ground state. However, the discontinuity in the momentum distribution cannot be interpreted in the usual way as evidence of perturbatively defined quasiparticles. The existence of quasiparticles is equivalent to the statement that the single-particle Green's function has poles with nontrivial residues (i.e., it possesses a quasiparticle weight $Z < 1$). According to Migdal's theorem, this form of the single-particle Green's function implies a discontinuity in the momentum distribution of the bare particles. However, the reverse is not necessarily correct: having a discontinuity in the momentum distribution does not imply the presence of a nontrivial residue in the single-particle Green's function and, therefore, the existence of quasiparticles. This is a necessary but not a sufficient condition for the presence of quasiparticles in the excitation spectrum of the system. Indeed, the discussion in Ref.~\onlinecite{Rozhkov} and its Supplemental Material indicates that, for qFL, the quasiparticle weight $Z$ vanishes for hole-like single-particle excitations at any finite momentum. This implies the lack of \emph{fermionic hole-like} quasiparticles in the low-lying excited states and thus, the state with $K=1$ is not merely a 1D FL.

\subsection{Spectral function}

An important quantity that can provide crucial information about the nature of the excitations in our 1D system is the single-particle spectral function:
\begin{align}
A(k,\omega) &= A_p(k, \omega)+A_h(k, \omega), \\
A_p(k, \omega) &= \sum_\alpha |\langle \alpha|c^\dagger_k|0\rangle|^2\delta(\omega-E_\alpha+E_0), \\
A_h(k, \omega) &= \sum_\alpha |\langle \alpha|c_k|0\rangle|^2\delta(\omega+E_\alpha-E_0), 
\label{Akw}
\end{align}
which contains contributions from both, the particle $A_p(k, \omega)$ and the hole $A_h(k, \omega)$ spectral functions. In these expressions $|0\rangle$ is the ground state of the system with $N$ particles and ground state energy $E_0$, and the states $|\alpha\rangle$ are excited states with one extra particle or hole, $N \pm 1$, and energy $E_\alpha$. In a free theory, the spectral function is given by a Dirac delta $A_p(k,\omega)\sim \delta(\omega-\epsilon_k)$, where $\epsilon_k$ is the dispersion relation. In the presence of interactions, FL theory dictates that the band curvature will be renormalized and peaks will now be broadened into Lorentzians, with a width determined by the quasiparticle lifetime. However, in 1D this picture breaks down due to the pervasive nesting and, instead, the low-energy physics is described by LL theory. In the particular soluble case of a linear dispersion, there is a particle-hole symmetry that is preserved and the excitation spectrum is a continuum with a power-law ``edge singularity'' given by
\begin{equation}
A(k,\omega)\sim \frac{\gamma_0^2}{(\omega-\epsilon_k)^{1-\gamma_0^2}}\theta[(\omega-\epsilon_k)\mathrm{sgn}\,\epsilon_k],
\end{equation}
where $\gamma_0$ is a constant that depends on the interaction potential~\cite{Dzya1974, Luther1974, Meden1992, Voit1993, Khodas2007}. Notice that, while the particle (hole) spectrum diverges as one approaches the mass shell $\omega=\epsilon(k)$ from above (below), it also displays a sharp threshold or discontinuity on the opposite side. As previously pointed out, the presence of curvature in the dispersion $\epsilon_k$ or irrelevant interactions produce some quite dramatic effects~\cite{Khodas2007, Imambekov2009, Imambekov2009b, Pereira2009, Imambekov2012, Markhof2016}: due to the broken particle-hole symmetry, the spectral function in the particle sector, near $k_F$, now displays a Lorentzian peak near the mass shell $\omega=\epsilon_k$, a feature of a FL; away from $\epsilon_k$, it displays an asymmetric two-sided power law. Notably, the spectral function in the hole sector still preserves the character of a LL.

\begin{figure}%[ht]
\centering
\includegraphics[width=0.48 \textwidth]{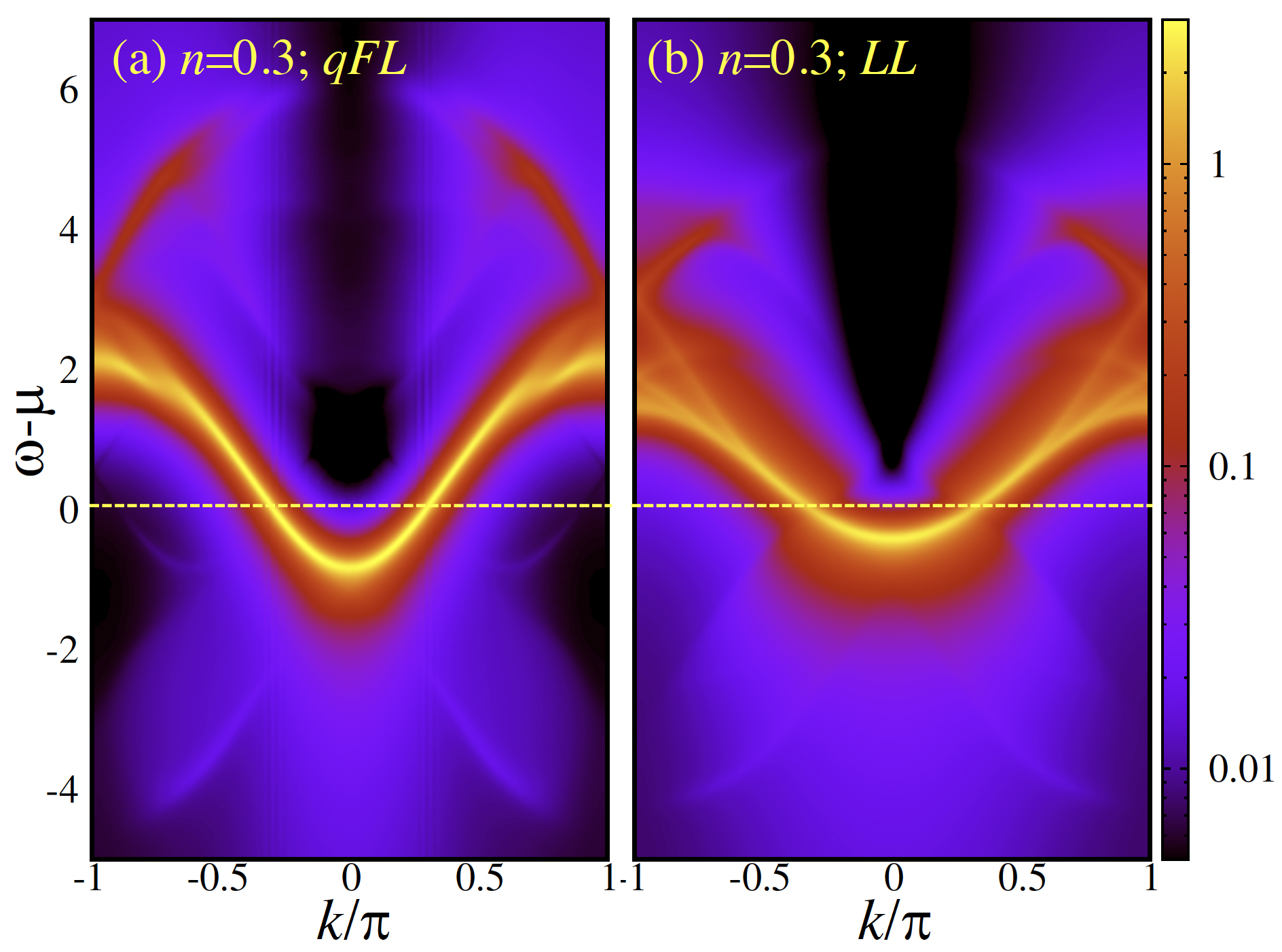}
\caption{Momentum resolved spectral function at density $n=0.3$ in a logarithmic color scale for (a) a quasi-Fermi liquid with $(V,V_2)=(-1,1)$ and (b) a Luttinger liquid with $(V,V_2)=(-1.5,0)$. Dashed lines indicate the Fermi level, where $k_F / \pi = 0.3$. Notice the renormalization of the bandwidth for the quasi-Fermi liquid candidate, the particle-hole asymmetry, the faint continuum in the hole spectrum ($\omega < \mu$) and the high energy bound state on the particle side ($\omega > \mu$).
\label{fig:akw}
}
\end{figure}

In order to characterize the spectrum of our model we carry out large-scale tDMRG simulations on 1D chains.  We consider systems with up to $L=240$ sites using $m=400$ DMRG states and a time step $\delta t=0.05$. This guarantees a truncation error smaller than $10^{-7}$ for times up to $t_{\max}=80$ (this implies that the main source of error stems from the Suzuki-Trotter decomposition). We follow the prescription detailed in Ref.~\onlinecite{Pereira2009} with minor modifications: to account for the open boundary conditions, we use a Hann window in real space that damps the effects of the edges and an exponential Hann window in time $w(t)=\exp{(-\epsilon t)}$, such that a Dirac delta peak would now have a Lorentzian line shape with an artificial broadening $\epsilon$,
\begin{equation}
A(\omega) \sim \frac{1}{\pi}\frac{\epsilon}{\omega^2+\epsilon^2}.
\end{equation}
As long as the inter-level spacing is much smaller than $\epsilon$ we should not expect noticeable finite-size effects~\cite{Jeckelmann2002, Paech2014}. If we assume that the bandwidth is $W\sim 4t$, the level spacing is of the order of $2W/L\sim 8t/L$. 

\begin{figure}%[ht]
\centering
\includegraphics[width=0.48 \textwidth]{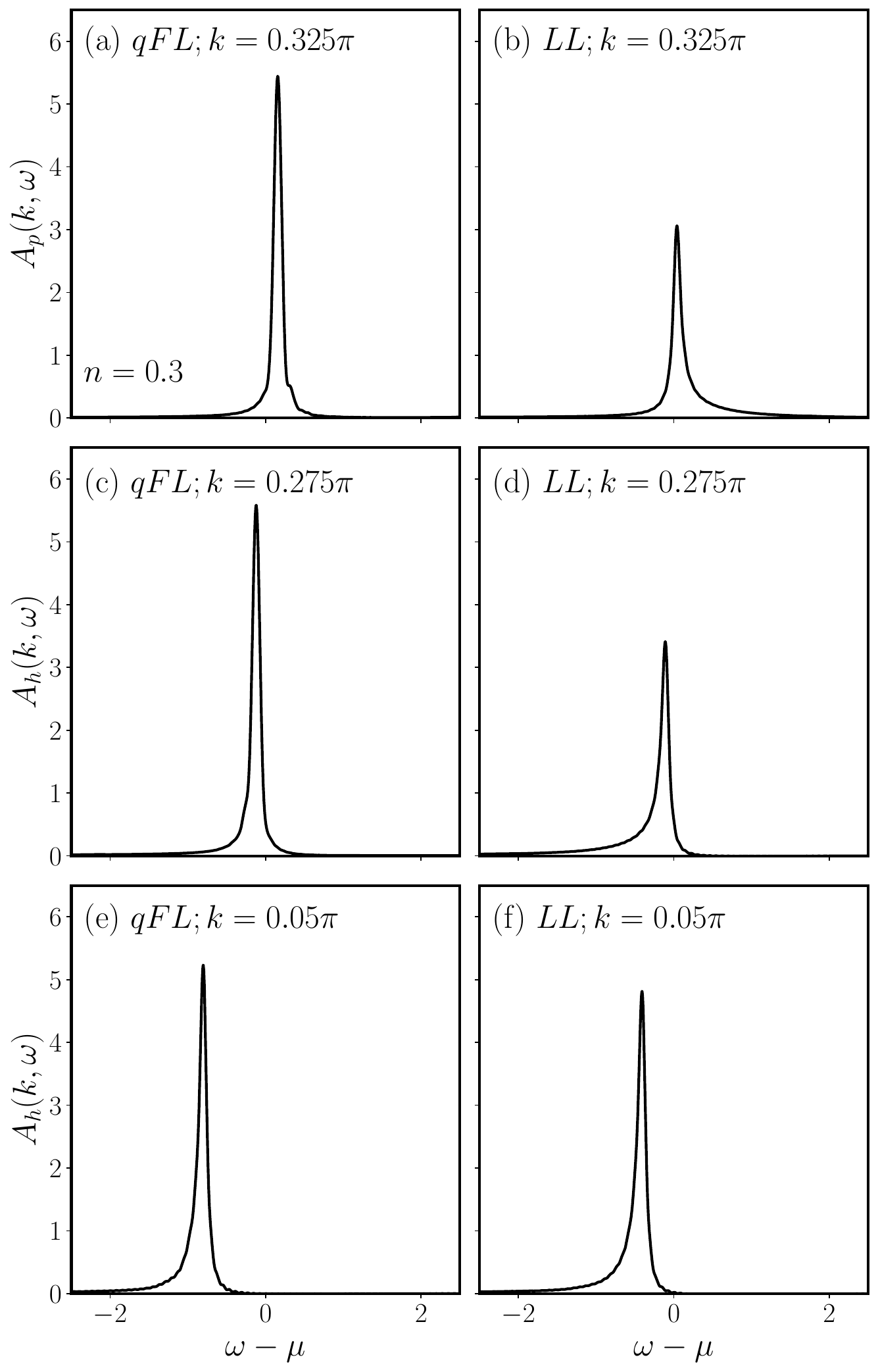}
\caption{Momentum cuts for the particle (a), (b) and hole (c)-(f) sectors of the spectral function at density $n=0.3$, for several values of momentum near and away from the Fermi points, located at $k_F=0.3\pi$. Left columns correspond to a quasi-Fermi liquid with $(V,V_2)=(-1,1)$ and right columns, to a Luttinger liquid with $(V,V_2)=(-1.5,0)$. Results are obtained with an artificial broadening $\epsilon=0.05$.
\label{fig:cuts}
}
\end{figure}

We focus our study on the regime with density $n=0.3$ with $(V,V_2) = (-1,1)$, such that the band curvature $d^2 \epsilon_k/dk^2 > 0$ at the Fermi level is more noticeable. To aid intuition, we first show the full momentum-resolved spectral function in Fig.~\ref{fig:akw}, compared to that of a LL with the same density and $(V,V_2)=(-1.5,0)$. The logarithmic scale allows us to clearly resolve the asymmetry between the particle ($\omega > \mu$) and hole ($\omega < \mu$) sectors, with an obvious broadening that increases with the distance from $k_F$ and a high-energy branch near $k=\pi$ that can be associated to bound states~\cite{Pereira2009}. Besides the obvious change in the effective mass and the bandwidth, the qFL may seem to possess a more coherent dispersion near $k_F$, as predicted using field-theory methods~\cite{Rozhkov2006, Rozhkov}. Regardless, for the hole spectrum we can observe a faint continuum that spreads to high energies in both cases shown.

\begin{figure}%[ht]
\centering
\includegraphics[width=0.48 \textwidth]{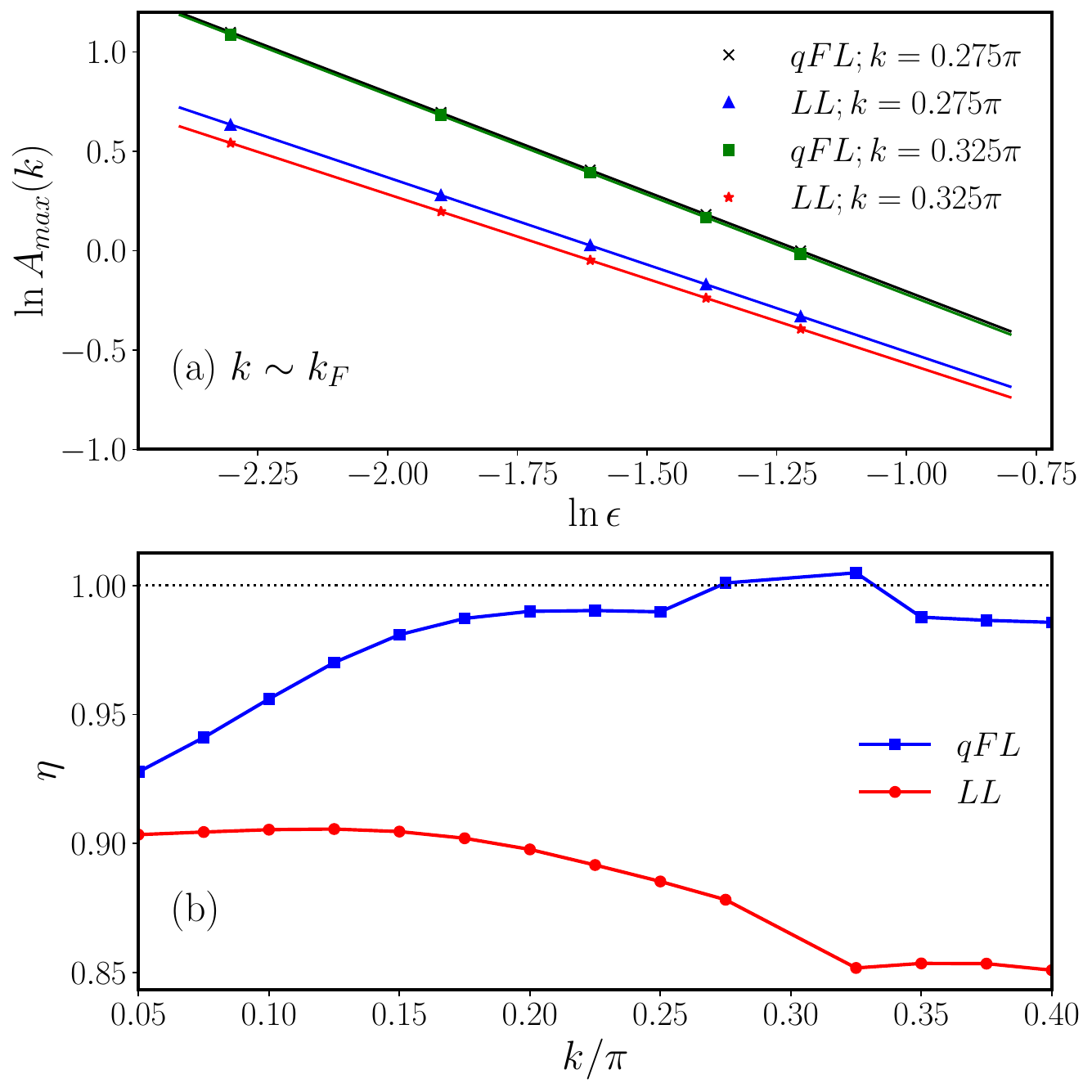}
\caption{(a) Scaling of the maximum in the spectral function for fixed values of $k$ near $k_F$, for the particle ($k > k_F = 0.3\pi$) and hole ($k < k_F = 0.3\pi$) sectors, both for the quasi-Fermi liquid and the Luttinger liquid having density $n=0.3$. (b) Power-law exponent obtained from the slopes in (a) for different values of momentum $k$.
\label{fig:peaks}
}
\end{figure}

We carry our study of the line shape near, and away from, the Fermi points as a function of the broadening $\epsilon$. In Fig.~\ref{fig:cuts} we show several cuts at fixed values of momenta for $\epsilon=0.05$ (we point out that we avoid sitting exactly at $k=k_F$ because this is the ``edge'' of the particle/hole band and due to the open boundary conditions and the Hann window used in the Fourier transform, results may be more affected). For the LL we observe a clear asymmetry resembling an edge singularity, while a similar asymmetry is evident for the qFL only at high energies, near the bottom of the band. The sharp peak-like structures displayed by the qFL at momenta $k \sim k_F$ may suggest the possibility of well-defined fermionic quasiparticles near the Fermi level. To extract more conclusive evidence, we first look at the scaling of the peak maximum $A_{\max}$ as a function of $1/\epsilon$, as shown in Fig.~\ref{fig:peaks}(a) in a logarithmic scale. Following Ref.~\onlinecite{Benthien2004}, a fit to a power law $A_{\max}\sim \epsilon^{-\eta}$ yields exponents $\eta=1.001$ and $\eta=1.004$ for $k=0.325\pi$ and $k=0.285\pi$, respectively, a strong indication that the peaks may be Lorentzians that evolve toward a Dirac delta in the limit $\epsilon \rightarrow 0$. These values give an interpolated $\eta \approx 1.002$ at $k_F$, suggesting an accuracy of at least $0.2\%$, taking into account that the values of $V$ and $V_2$ had to be fine-tuned and there is an error stemming from this estimate as well. Notice that we avoid values of $k$ too close to $k_F$ to avoid artifacts created by the open boundary conditions and the Hann window~\cite{Pereira2009}. This is in agreement with the fact that $K \approx 1$. In contrast, the results for the LL show that $\eta$ differs noticeably from unity, and from each other, as expected from a power-law singularity edge. To validate this approach, we also fit the exponent for a LL at half filling with $(V,V_2)=(-1.5,0)$: the Bethe-ansatz prediction is $\eta=0.8416$~\cite{Pereira2009} while we obtain $\eta=0.8262$ for $k=0.475\pi$. In Fig.~\ref{fig:peaks}(b) we show the power-law exponent $\eta$ as a function of momentum $k$ obtained from the fits. Remarkably, for the qFL we see a clear change of behavior near the Fermi momentum, with the exponent approaching values very close to $\eta=1$, suggesting that the FL-like behavior occurs only near the Fermi level. %, in agreement with having $K = 1$ (free fermions). 

\begin{figure}%[ht]
\centering
\includegraphics[width=0.48 \textwidth]{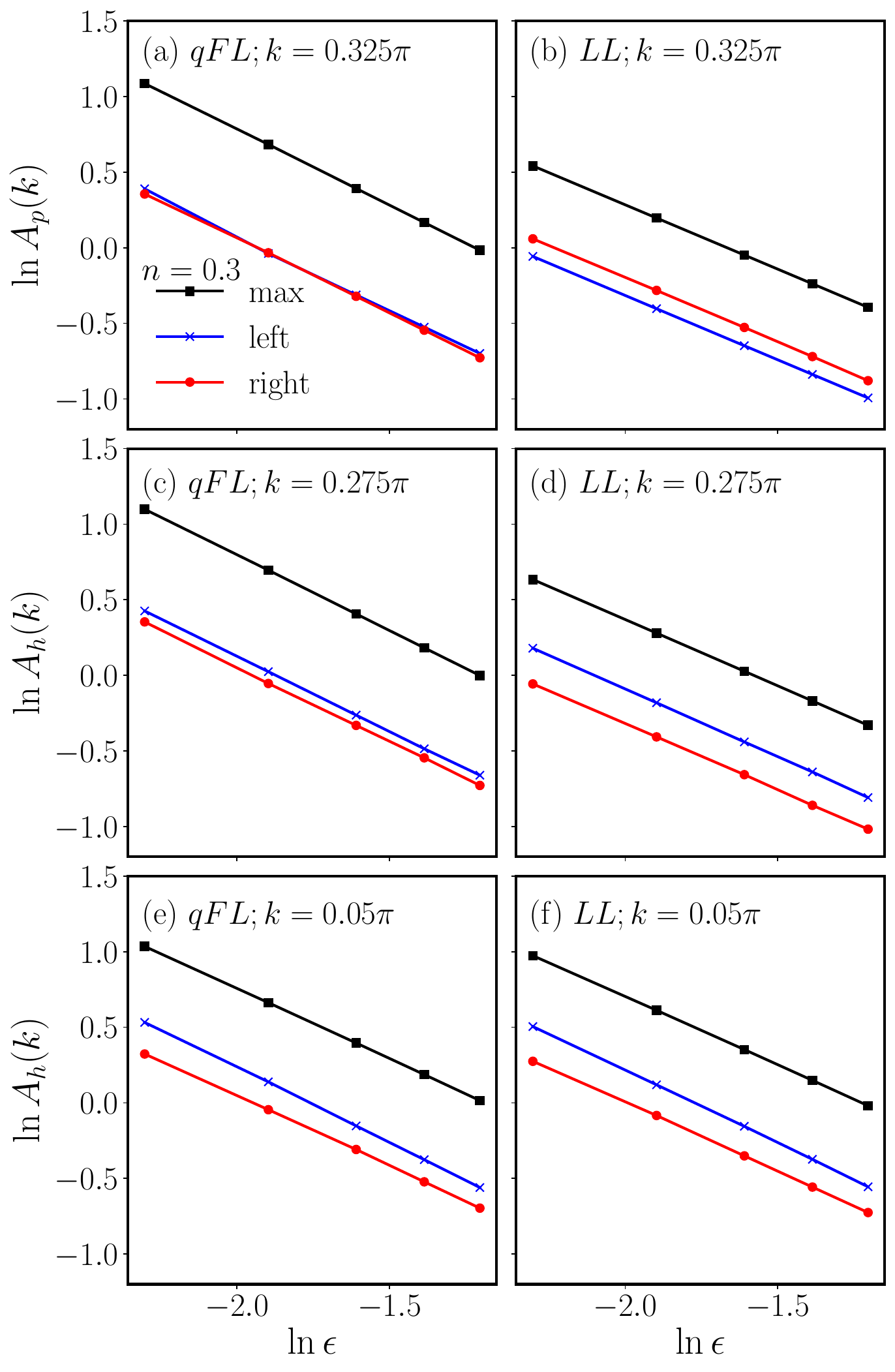}
\caption{Scaling of the maximum $A_{\max}$ and the value of the spectral function at $\omega_{\max}\pm \epsilon$ for both the quasi-Fermi liquid and Luttinger liquid at density $n=0.3$. We show results on the particle ($k > 0.3\pi$) and hole ($k < 0.3\pi$) sectors of the spectrum for some representative momenta $k$. Lines connecting the data points are a guide to the eye. The same scaling and symmetry in panel (a) are a strong indication of quasiparticle behavior.
\label{fig:left_right}
}
\end{figure}

\begin{figure}%[ht]
\centering
\includegraphics[width=0.48 \textwidth]{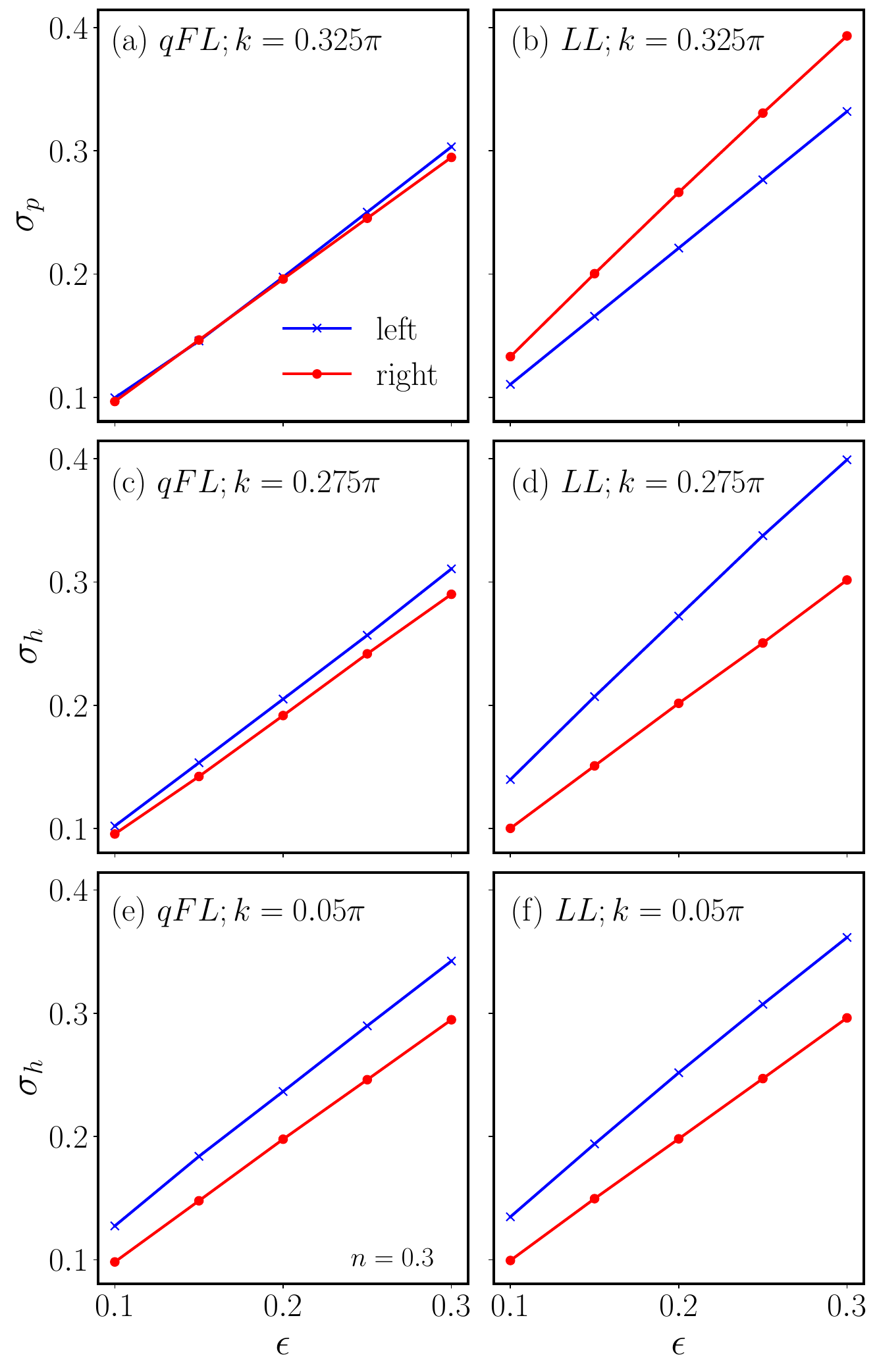}
\caption{Distance of the half maximum $A_{\max}/2$ from the position of the peak at $\omega_{\max}$, for both the quasi-Fermi liquid and Luttinger liquid, at density $n=0.3$. Lines connecting the data points are a guide to the eye. The linear scaling with slope one and the left/right symmetry in panel (a) are strong indications of quasiparticle behavior.
\label{fig:sigma}
}
\end{figure}

To further characterize the singularities in the spectra, we plot the weight of the line shape at $\omega_{\max}\pm \epsilon$, where $\omega_{\max}$ is the position of the maximum, $A_{\max}$. If the scaling is the same as that of $A_{\max}$, and there is no left/right asymmetry, the evidence in favor of a Dirac delta pole is strengthened. As seen in Fig.~\ref{fig:left_right}, where we show representative data, this occurs on the particle sector of the spectrum, although the evidence weakens on the hole side due to an apparent asymmetry that grows with distance from $k_F$. This asymmetry, more dramatic for small $k$, is already present in the LL for all values of $k$, a hallmark of an edge singularity. To corroborate this behavior we also show in Fig.~\ref{fig:sigma} the half width of the peak, measured to the left and to the right of the maximum $\sigma_\pm =|\omega_{\pm}-\omega_{\max}|$, where $\omega_\pm$ is the value of frequency at half maximum, $A(k,\omega_\pm)=A_{\max}/2$ from above and below. For an ideal Lorentzian, the points should lie on a straight line with slope 1, as it occurs in panel (a) for the qFL for $k \gtrsim k_F$. Once again, this points to the presence of a Dirac delta peak in the particle spectral function, but an edge singularity in the hole sector that broadens with an exponent $\eta$ that decreases as one moves away from $k_F$.

If indeed the system realizes full fledged Landau quasiparticles, one should be able to associate the magnitude of the discontinuity in $n(k)$ with the quasiparticle weight $Z=|\langle N+1|c^\dagger_k|0\rangle|^2$. Due to the open boundary conditions, we use ``particle-in-a-box'' states~\cite{Benthien2004}
\begin{equation}
c^\dagger_k = \sqrt{\frac{2}{L+1}}\sum_\ell \sin{(k \ell)} c^\dagger_\ell,
\end{equation}
with $k=\pi \ell/(L+1)$, for integers $1 \le \ell \le L$. An extrapolated value of this quantity to the thermodynamic limit yields a value $Z=0.945$ (not shown), which can be compared to the value of the pole in the spectral function at $k=0.325\pi$, $Z=0.940$. Conversely, the value of the discontinuity in Fig.~\ref{fig:momfes}(a) is 0.953, sufficiently close to support our conclusion.

\section{Summary and conclusion\label{sec:conclusions}}

We have presented numerical evidence for the existence of the qFL state in 1D, which departs from both the standard LL and FL phenomenologies, and yet combines features from each. We achieved this in a spinless fermion lattice Hamiltonian system, by effectively nullifying the marginal interactions such that the remaining \emph{irrelevant} interactions manifest in the state. We observed this for Hamiltonian parameters well beyond the perturbative regime, demonstrating the stability and legitimacy of the quasi-Fermi liquid state as a distinct metallic state of matter in one spatial dimension.

The continuum limit of our lattice model with generic $V$ and $V_2$ interactions corresponds to an effective low-energy theory with marginal density-density interactions and additional irrelevant terms. From there, one may elect to drop these irrelevant terms leading to the typical linear LL model. Notably, what we have effectively done is nullify the marginal interaction while keeping the irrelevant terms. In our lattice Hamiltonian we have accomplished this by judiciously selecting values of $(V, V_2, \mu)$, such that $K = 1$. This should account for retaining the irrelevant terms but ``dropping'' the marginal interactions. According to Ref.~\onlinecite{Rozhkov}, the qFL paradigm states that irrelevant modifications to the noninteracting theory qualitatively change the nature of the ground state and excitations in the system.

Specifically, the qFL ground state is perturbatively connected to the free-fermion ground state as for the FL. This is evident from $n(k)$, shown in Fig.~\ref{fig:momfes}, due to the finite discontinuity in $n(k = k_F)$ and the renormalization of the occupations: $n(k \lesssim k_F) < 1$ and $n(k \gtrsim k_F) > 0$. As for excited states, certainly, the divergent quasiparticle residue $Z$ in the hole sector implies that the excited states of a qFL are qualitatively different compared to those of free fermions. 
%At the level of the lattice, our meaningful deviation from the free Fermi gas is general lattice effects which include both band curvature and irrelevant interactions. 

As for excited states, the evidence presented here suggests that the spectral characteristics of the qFL should partly mimic those of the LL, but with well-defined fermionic quasiparticles for the addition spectrum $A_p$ near the Fermi momentum $k_F$. In this regime the particle-like excitations resemble FL quasiparticles. On the other hand, the removal spectrum $A_h$ displays edge singularities, which are asymptotically close to a Dirac delta as $k$ approaches $k_F$, signaling the absence of hole-like quasiparticles. These collective excitations are characteristic of LL. See Figs.~\ref{fig:cuts}-\ref{fig:sigma}.

{The role of the particle and hole sectors for $A(k, \omega)$ can be swapped via a particle-hole transformation. Therefore, it is possible to realize quasiparticles for $|k| < k_F$ and edge singularities for $|k| > k_F$. In this work, we have studied the case when $k_F < \pi / 2~(\mu > 0)$. Conversely, we can understand the $k_F > \pi / 2~(\mu < 0)$ case by exchanging particles and holes.}

%The need for fine-tuning the marginal interactions can be additionally understood in terms of RG flow and nontrivial effects from \textit{dangerously} irrelevant operators. Dangerously irrelevant operators introduce perturbations to the low-energy Hamiltonian that may affect the physics in the IR and lead to different fixed points in RG space. In other words, they make the RG flow chaotically sensitive to the initial parameter values. Since the scaling dimension of marginal interactions is higher than any irrelevant operator ($0$ vs. $< 0$), their effects on correlation functions will dominate; typically leading to the LL phase in 1D. 
The need for fine-tuning the marginal interactions originates from the fact that, without such a dominating term, the system stabilizes (flows to) a new phase: the qFL. This is how we understand the results of Fig.~\ref{fig:phasediag}: above or below the $K = 1$ line in the phase diagram, marginal interactions dominate yielding either a repulsive, charge-ordered phase $(K < 1)$ where $V_2$ dominates; or an attractive, superconducting-like phase $(K > 1)$ where $V$ dominates. 
%The presence of marginal interactions in either case causes the RG flow toward the typical 1D phases. 
For the $K = 1$ line, marginal interactions are functionally not present and the irrelevant terms secure the qFL phase, away from the low-energy perturbative regime.

%The evidence presented here suggests that the spectral characteristics of the qFL should mimic those of the nonlinear LL, but with well-defined fermionic quasiparticles for the addition spectrum $A_p$ near the Fermi momentum $k_F$. On the other hand, the removal spectrum $A_h$ displays edge singularities, which are asymptotically close to a Dirac delta as $k$ approaches $k_F$. %However, this is unclear since irrelevant terms to the linear LL are understood as deviations around the LL fixed point. 

%\section{Conclusion}

{ 
There are other contexts where the qFL can be realized, apart from the competition between nearest- and next-nearest-neighbor interactions $(V \approx -V_2)$, discussed in this work, %Indeed, although our results indicate that if a system presents frustration constraints the qFL can be stabilized, we note that more general lattice models can also display qFL behavior. 
for example, in Hamiltonian models that include next-nearest-neighbor and correlated hopping terms. Note, also, that the constraint $V \approx -V_2$ is not necessary for the stabilization of the state~\cite{suppmat}. Hence we hypothesize that lattice Hamiltonians with general quartic interactions, beyond the nullification of marginal terms, are candidates to exhibit a qFL phase~\footnote{J. D. Baktay \emph{et al.}, [in preparation].}. These conclusions might be extended to spinful fermions~\cite{Essler2015}.} 
%As a matter of fact, the Hamiltonian studied here manifests such a constraint as $V \approx -V_2$ whenever the qFL is realized. %First, we found similar results for $\mu=0 (n=0.5)$ which can be seen in the Supplemental Material. 
%As described in Ref.~\onlinecite{Rozhkov}, the qFL can be achieved without fine-tuning by adding irrelevant terms explicitly to the Hamiltonian. These terms can either be quadratic corrections to the relativistic dispersion relation or density-current interactions. 
%Hence, we hypothesize other candidate Hamiltonians that might include next-nearest neighbor hopping or correlated hopping terms.

{Lastly, we would like to note that a strict fine-tuning of the interactions is not necessary. As discussed in Sec.~\ref{sec:results}, satisfying $|K - 1| < \delta$, where $\delta \ll 1$, is enough to detect the qFL signatures in the momentum distribution and the spectral function. Thus, an experimental realization could potentially not be as demanding. For instance, the qFL may be realized in optical lattice setups of fully polarized fermions, where fine control of interactions can systematically be engineered~\cite{Esslinger2010, Schafer2020}.}

\begin{acknowledgments}
AEF and JDB acknowledge support from the U.S. Department of Energy, Office of Basic Energy Sciences under grant No.\ DE-SC0014407 (AEF and JDB). JR acknowledges support from the Office of the Vice-president of Research and Creative Activities and the Office of the Faculty of Science Vice-president of Research of Universidad de los Andes under the FAPA grant. JDB thanks P.\ Weinberg for valuable interactions at the initial stages of developing the MPS code.
\end{acknowledgments}

%\nocite{*}
%\section{References}
%\bibliographystyle{apsrev4-2}

\bibliography{references}

\end{document}

\title{Supplemental Material for ``Quasi-Fermi liquid behavior in a one-dimensional system of interacting spinless fermions"}

\author{Joshua D. Baktay}
\affiliation{Department of Physics, Northeastern University, Boston, Massachusetts 02115, USA}

\author{Alexander V. Rozhkov}
\affiliation{Institute for Theoretical and Applied Electrodynamics, Russian Academy of Sciences, 125412 Moscow, Russia}

\author{Adrian E. Feiguin}
\affiliation{Department of Physics, Northeastern University, Boston, Massachusetts 02115, USA}

\author{Juli\'an Rinc\'on}
\affiliation{Department of Physics, Universidad de los Andes, Bogot\'a D.C. 111711, Colombia}

\date{\today}

\maketitle

\section{Introduction}

In the main text we characterized the ground and excited states of a quasi-Fermi liquid (qFL) system. For the ground state, we saw FL-like behavior in accordance with Ref.~\onlinecite{Rozhkov}. For the excited states we studied the line shape of the single-particle spectral function at various momenta. There, we found evidence of well-defined fermionic quasiparticles in the addition spectrum near the Fermi momentum, $k_F$, and edge singularities in the removal spectrum for $k < k_F$. The former is characteristic of FL behavior while the latter is characteristic of Luttinger liquid (LL) behavior. Together these behaviors are generally suggestive of non-linear LL theory~\cite{Pereira2009, Periera2012, Imambekov2009, Imambekov2009b, Imambekov2012}. Here, we present additional evidence for the existence and stability of the qFL in different interaction coupling regimes. In Sec.~\ref{sec:section2} we present the other viable region of Hamiltonian parameter space for quasi-Fermi liquid behavior: $(V > 0, V_2 < 0)$. In Sec.~\ref{sec:section3} we present additional scaling analyses for the quasi-Fermi liquid candidates highlighted in the main text.

\begin{figure}
    \includegraphics[width=.43\textwidth]{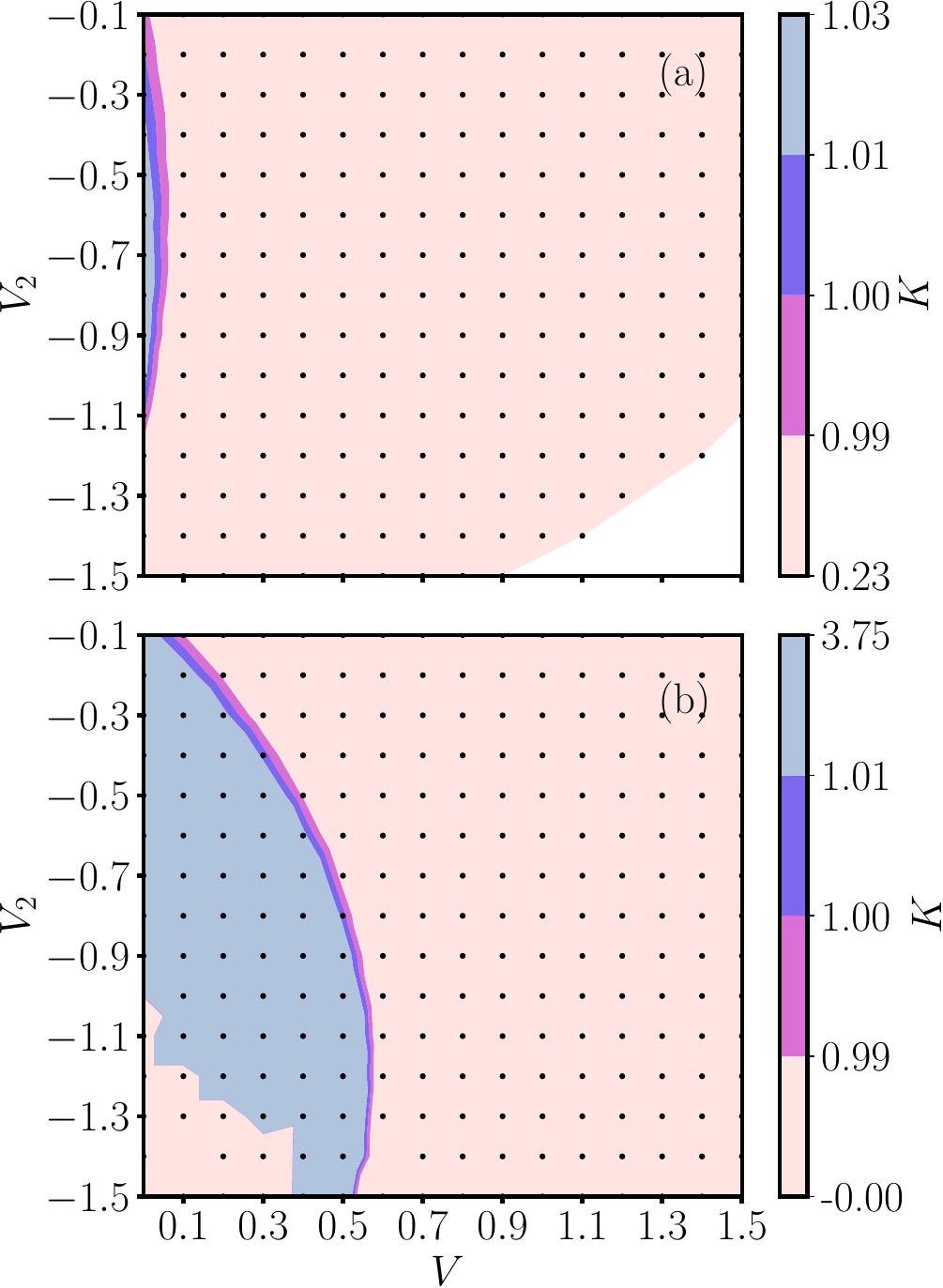}
    \caption{Phase diagram of Luttinger parameter, $K$, for $V > 0$ and $V_2 < 0$ at $\mu/t = +0.5$ (a) and $\mu/t = +1.0$ (b). The qFL state is stable along the band defined by $0.99 < K < 1.01$. Note the qualitatively different behavior of the bands compared to those in the main text. Additionally, in the lower right hand quadrant of (a), the leading eigenvalue of the MPS transfer matrix is degenerate, signaling breaking of translation invariance preventing ground state convergence, thus resulting in the missing data points. In the lower left quadrant of (b), we have a charge-ordered state which also breaks translation invariance, yielding $K \approx 0$ or preventing convergence.}
    \label{fig:Kphase}
\end{figure}

\section{Additional quasi-Fermi liquid ground states \label{sec:section2}}

As described in the manuscript, since we expect the qFL ground state to be perturbatively connected to the non-interacting case, we searched for states with Luttinger parameters $K = 1$. In the main text we focused on the upper-left quadrant of Hamiltonian parameter space such that $V \cdot V_2 < 0$, which forces competition between repulsive (positive) and attractive (negative) interactions. Here, we present the lower-right quadrant of Hamiltonian parameter space such that the same condition is satisfied, namely $(V > 0, V_2 < 0)$, for values of the chemical potential $\mu = 0.5, 1$. As in the main text, all energy coupling are reported in units of the hopping amplitude $t$.

In both plots of Fig.~\ref{fig:Kphase} we find an approximate $K = 1$ line of qFL candidates. However, in this quadrant $V$ and $V_2$ are no longer comparable in value. Although for larger values of $\mu$ that may indeed be the case.
%Previously, the nearest neighbor coulomb interactions were attractive $(V < 0)$ and the next nearest neighbor interactions were repulsive $(V_2 > 0)$. 
%In this quadrant, the attractive and repulsive interactions are swapped. 
The sign change in the interactions fundamentally changes the nature of the leading quantum order fluctuations, and therefore, the energy scales at which both marginal interactions (those associated to $V$ and $V_2$) can mutually nullify each other. This is particularly dramatic for $\mu = 0.5$, as shown in Fig.~\ref{fig:Kphase}(a). For $\mu = 1.0$, depicted in Fig.~\ref{fig:Kphase}(b), the lower particle number allows for a more subtle interplay between the attractive and repulsive forces necessary to stabilize the qFL phase.

\begin{figure}
    \centering
    \includegraphics[width=.4\textwidth]{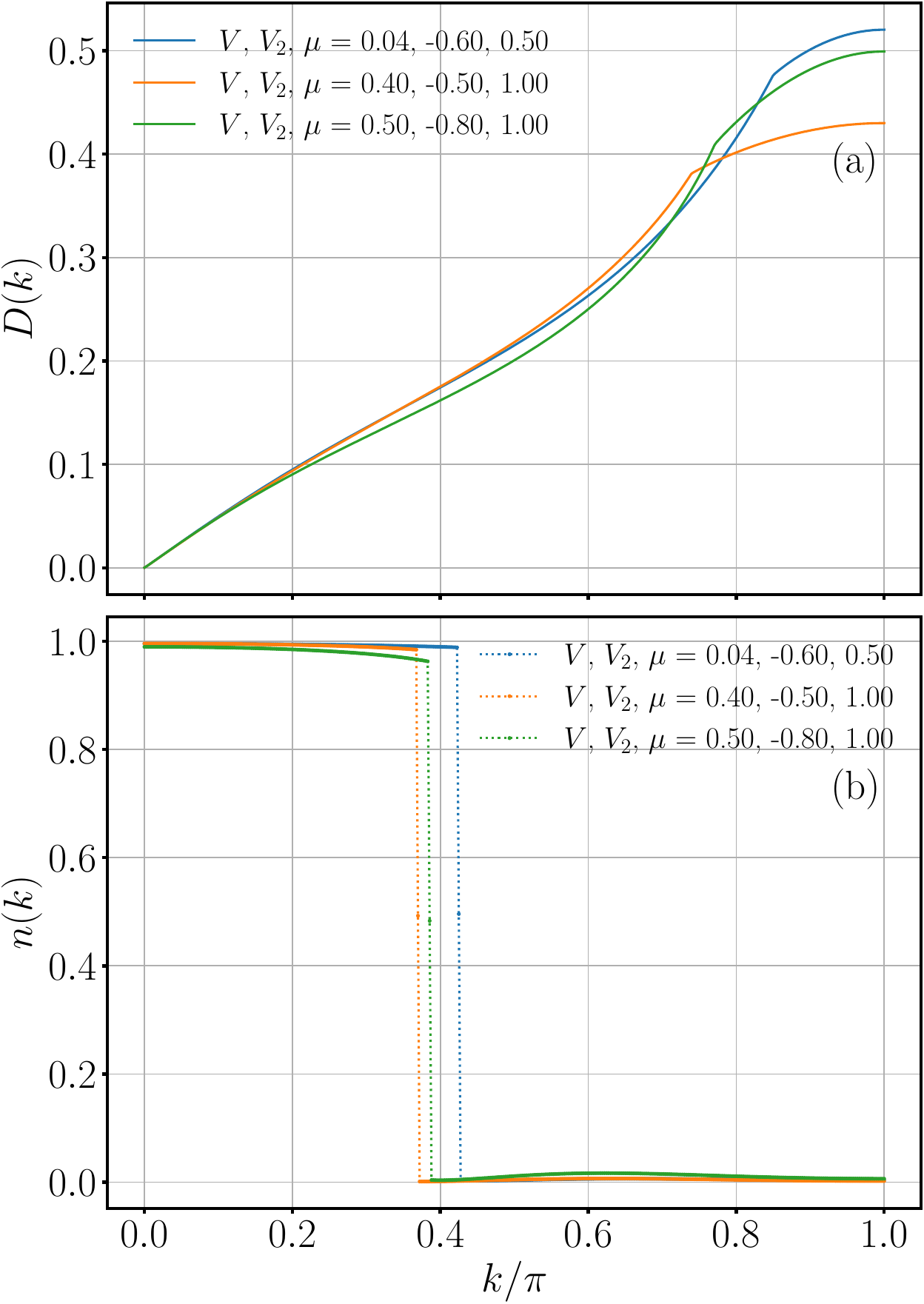}
    \caption{Charge structure factor, $D(k)$, (a) and momentum distribution, $n(k)$, (b) for three representatives of the qFL bands in Fig.~\ref{fig:Kphase}. In (a), we clearly see discontinuous behavior analogous to the main text. In (b), we see interacting behavior, departing from the plateau-like behavior for a free Fermi gas. From $n(k)$, we calculate $\Delta n(k_F)$ and from $D(k)$ we calculate $K$ from a linear, low-momentum approximation. See Fig.~\ref{fig:delta}.
    }
    \label{fig:scfs}
\end{figure}

Nevertheless, these qFL candidates exhibit similar characteristics to those shown in the main text. This can be seen via three representative examples shown in Fig.~\ref{fig:scfs}. Note that the momentum distribution functions behave identically to those in the upper-left quadrant (specifically, the renormalizations of the occupations for $k < k_F$ and $k > k_F$); see Fig.~\ref{fig:scfs}(b). However, the charge static structure factors are qualitatively different for $k>2k_F$, even though they still demonstrate interacting, metallic behavior and the absence of charge order. This is depicted in Fig.~\ref{fig:scfs}(a). As expected, as the couplings become weaker the resemblance to the non-interacting case becomes stronger. The most important feature of $n(k)$ is the sharp, robust, finite discontinuity distinctive of a FL.

\section{Additional Scaling Analyses\label{sec:section3}}

\subsection{Finite-entanglement Scaling}

\begin{figure}
    \centering
    \includegraphics[width=.44\textwidth]{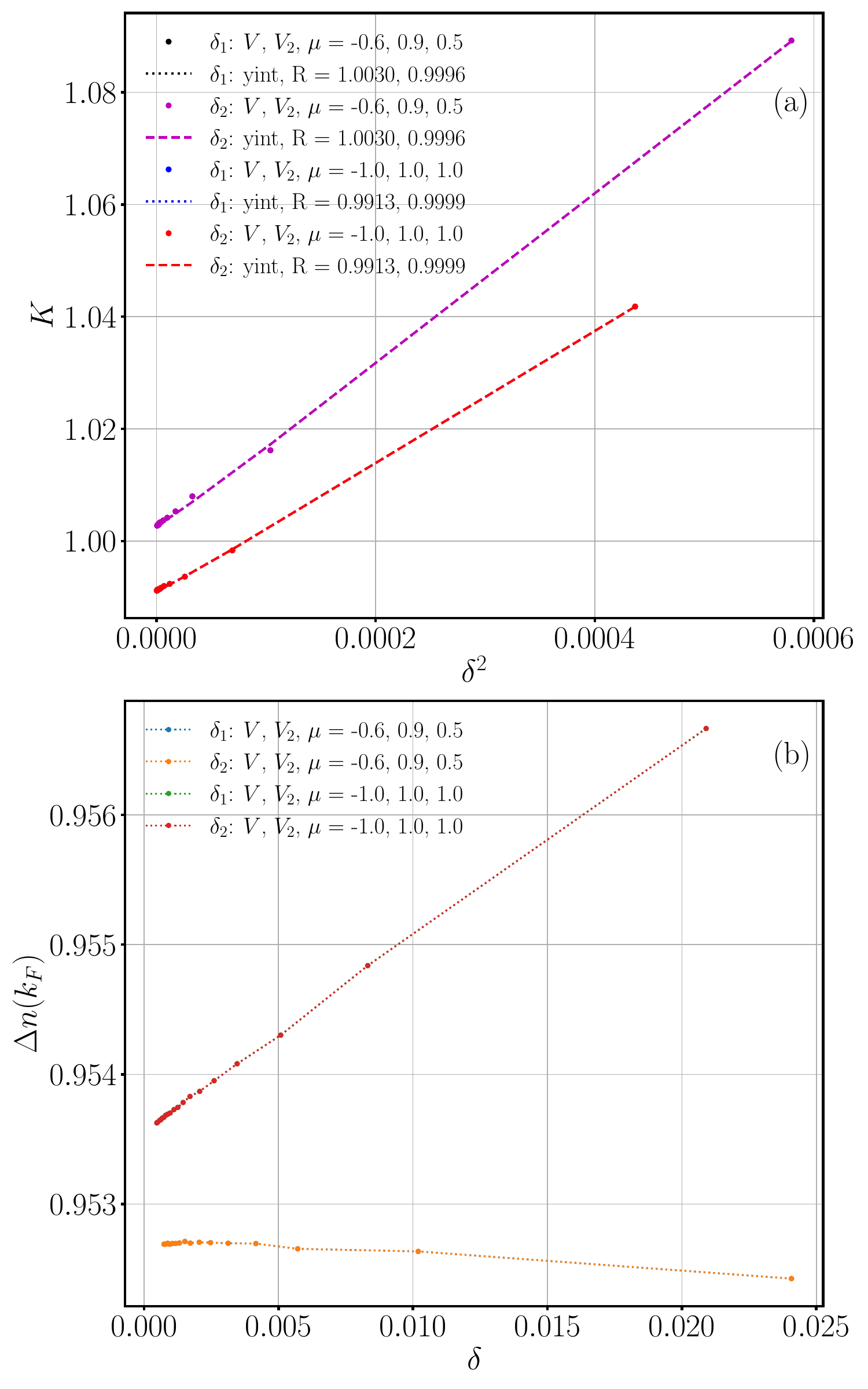}
    \caption{Luttinger parameter, $K$, (a) and size of the discontinuity, $\Delta n(k_F)$, (b) as a function of two equivalent finite-entanglement length scales, $\delta \in \{\delta_1, \delta_2\}$. In (a), the intercepts of quadratic fits yield $K \approx 1$ in the $\delta \to 0$ limit which corresponds to $\xi \to \infty$. In (b), as $\delta \to 0$, $\Delta n(k_F)$ converges to a finite value $\approx 0.95$ in agreement with the main text and further demonstrates the presence of a discontinuity. Lines are added to guide the eye. Note, in both plots, not all curves are visible due to the equivalence between $\delta_1$ and $\delta_2$. 
    }
    \label{fig:delta}
\end{figure}

In this section we present additional characterizations of the qFL ground states presented in the main text. As discussed there, we conducted a fininte-entanglement scaling analysis to determine the Luttinger parameter $K$ in the infinite-bond dimension limit, by studying the scaling of $K$ versus the correlation length of the MPS. As alluded to in the main text, here we present a similar analysis using the length scale $\delta = \xi_n^{-1} - \xi_1^{-1}$, in place of $\xi_1(\chi)$, where $\xi_n^{-1} = -\log|\lambda_{n}(\chi)|$ and $\lambda_{n}(\chi)$ is the $n$-th sub-leading eigenvalue of the matrix product state (MPS) transfer matrix~\cite{Cincio2018}. As shown in Fig.~\ref{fig:delta}(a) we find a quadratic scaling behavior for two equivalent definitions: $\delta_1 \equiv \xi_2^{-1} - \xi_1^{-1}$ and $\delta_2 \equiv \xi_3^{-1} - \xi_1^{-1}$. As $\xi \to \infty$, $\delta \to 0$ and all four curves converge to $K \approx 1$ again with a statistical correlation $R \simeq 1$. Note, that not all four curves $(\delta_1$ and $\delta_2$ for each parameter set$)$ are visible in Fig.~\ref{fig:delta} since $\delta_1$ and $\delta_2$ lie nearly perfectly on top of each other. This is numerical confirmation of the equivalence between the possible length scales, $\delta$, and the MPS correlation length, as originally shown in Ref.~\onlinecite{Cincio2018}.

In Fig.~\ref{fig:delta}(b) we show the same analysis for the extrapolation of the magnitude of the discontinuity about $k_F$. Again, the lines clearly extrapolate to a finite value in agreement with the scaling of the correlation length.

%All together, this gives a robust characterization of the qFL ground state. 

\subsection{Ground-state entanglement entropy}

Finally, we present the scaling of the entanglement entropy of the ground state of the qFL, $S$, as a function of the correlation length $\xi$.

With the approximation to the ground state in terms of a uMPS, it is straightforward to compute $S$ for a half-infinite chain from the Schmidt decomposition of the uMPS~\cite{schollwock2011, lecture_notes}. For a critical system, it has been shown using conformal field theory that $S$ scales logarithmically with the correlation length as~\cite{Wilczek, Vidal2003, Calabrese2004, FES}:
\begin{equation}
S = \frac{c}{6} \log (\xi / a)
\end{equation}
where $c$ is the central charge of the underlying conformal field theory, and $a$ is the lattice spacing (set to 1, in our case). The central charge, which is a measure of the number of degrees of freedom in the system, corresponds to $c = 1$ for free real bosons and to $c = \frac{1}{2} + \frac{1}{2} = 1$ for free complex fermions. This signals the boson-fermion correspondence in two space-time dimensions that can be proved using bosonization~\cite{DiFrancesco, giamarchi, gogolinbook}.

\begin{figure}
    \includegraphics[width=.42\textwidth]{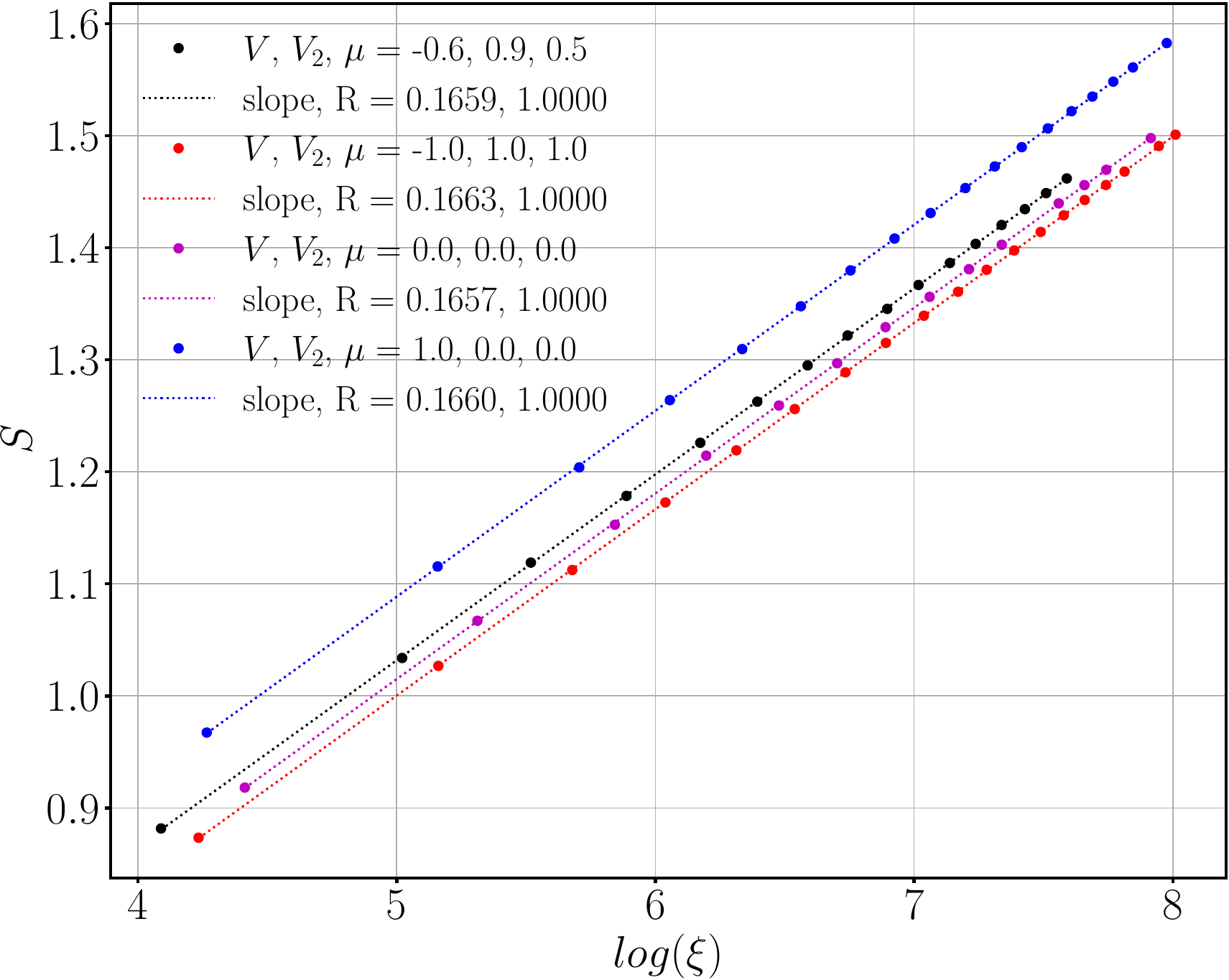}
    \caption{Entanglement entropy of the ground state, $S$, as a function of the correlation length, $\xi$, for two quasi-Fermi liquid representative candidates, the free Fermi gas, and a known Luttinger liquid. The logarithmic plot affords a linear fit whose slope can be compared to the analytical behavior (see text). As expected, the behavior of the different ground states are very similar, yielding a slope $ \approx \frac{1}{6}$. 
    \label{fig:Sfes}
    }
\end{figure}

Figure~\ref{fig:Sfes} shows the scaling behavior of the ground state entanglement entropy, $S$, for two qFL candidates, the free Fermi gas, and a known LL state. Fitting the results to the expression above, all of those results yield a central charge $c \approx 1$, with an approximate error of about $0.5\%$ (which, in passing, makes manifest the reliability of the uMPS as an accurate representation of the qFL ground state). These results imply that the low-energy effective field theory that describes the qFL phase corresponds to free complex fermions or free real bosons, according to the field theory predictions. Given that the spectral function results indicate that the excitations at $k_F$ are fermionic (see main text), the low-energy field theory may be best understood starting from a free complex fermion theory. This is indeed what has been previously discussed; see Refs.~\onlinecite{Periera2012, Imambekov2012, Rozhkov} and references therein.

\bibliographystyle{apsrev4-1}
\bibliography{references}